\documentclass[%
reprint,
superscriptaddress,
amsmath,amssymb,
aps,
prx
]{revtex4-2}
\usepackage[utf8]{inputenc}
\usepackage[english]{babel}
\usepackage{blindtext}
\usepackage{graphicx}   
\graphicspath{{figs/}}   
\usepackage{dcolumn}
\usepackage{bm}
\usepackage[hidelinks]{hyperref}

\usepackage{epstopdf}

\usepackage{braket}
\usepackage{amsmath}
\usepackage{siunitx}
\usepackage{changes}
\usepackage{natbib}
\usepackage[title]{appendix}
\usepackage{multirow}

\begin{document}


\title{Implementation of a transmon qubit using superconducting granular aluminum}

\author{Patrick Winkel}
\affiliation{Physikalisches Institut, Karlsruhe Institute of Technology, 76131 Karlsruhe, Germany}

\author{Kiril Borisov}
\affiliation{Institute of Nanotechnology, Karlsruhe Institute of Technology, 76344 Eggenstein-Leopoldshafen, Germany}

\author{Lukas Grünhaupt}
\affiliation{Physikalisches Institut, Karlsruhe Institute of Technology, 76131 Karlsruhe, Germany}

\author{Dennis Rieger}
\affiliation{Physikalisches Institut, Karlsruhe Institute of Technology, 76131 Karlsruhe, Germany}

\author{Martin Spiecker}
\affiliation{Physikalisches Institut, Karlsruhe Institute of Technology, 76131 Karlsruhe, Germany}

\author{Francesco Valenti}
\affiliation{Physikalisches Institut, Karlsruhe Institute of Technology, 76131 Karlsruhe, Germany}
\affiliation{Institut für Prozessdatenverarbeitung und Elektronik, Karlsruhe Institute of Technology, 76344 Eggenstein-Leopoldshafen, Germany}

\author{Alexey V. Ustinov}
\affiliation{Physikalisches Institut, Karlsruhe Institute of Technology, 76131 Karlsruhe, Germany}
\affiliation{Russian Quantum Center, National University of Science and Technology MISIS, 119049 Moscow, Russia}

\author{Wolfgang Wernsdorfer}
\affiliation{Physikalisches Institut, Karlsruhe Institute of Technology, 76131 Karlsruhe, Germany}
\affiliation{Institute of Nanotechnology, Karlsruhe Institute of Technology, 76344 Eggenstein-Leopoldshafen, Germany}
\affiliation{Institut Néel, CNRS and Université Joseph Fourier, Grenoble, France}

\author{Ioan M. Pop}
\email{ioan.pop@kit.edu}
\affiliation{Physikalisches Institut, Karlsruhe Institute of Technology, 76131 Karlsruhe, Germany}
\affiliation{Institute of Nanotechnology, Karlsruhe Institute of Technology, 76344 Eggenstein-Leopoldshafen, Germany}

\date{\today}

\begin{abstract}
The high kinetic inductance offered by granular aluminum (grAl) has recently been employed for linear inductors in superconducting high-impedance qubits and kinetic inductance detectors. Due to its large critical current density compared to typical Josephson junctions, its resilience to external magnetic fields, and its low dissipation, grAl may also provide a robust source of non-linearity for strongly driven quantum circuits, topological superconductivity, and hybrid systems. Having said that, can the grAl non-linearity be sufficient to build a qubit? Here we show that a small grAl volume ($10 \times 200 \times 500 \,\mathrm{nm^3}$) shunted by a thin film aluminum capacitor results in a microwave oscillator with anharmonicity $\alpha$ two orders of magnitude larger than its spectral linewidth $\Gamma_{01}$, effectively forming a transmon qubit. With increasing drive power, we observe several multi-photon transitions starting from the ground state, from which we extract $\alpha = 2 \pi \times 4.48\,\mathrm{MHz}$. Resonance fluorescence measurements of the  $\ket{0} \rightarrow \ket{1}$ transition yield an intrinsic qubit linewidth $\gamma = 2 \pi \times 10\,\mathrm{kHz}$, corresponding to a lifetime of $16\,\si{\micro\second}$, as confirmed by pulsed time-domain measurements. This linewidth remains below $2 \pi \times 150\,\mathrm{kHz}$ for in-plane magnetic fields up to $\sim70\,\mathrm{mT}$. 
\end{abstract}


\maketitle

Superconducting circuits are part of a growing group of hardware platforms which successfully demonstrated quantum information processing, from quantum error correction to quantum limited amplification \cite{Krantz19}. Some of the most promising platforms, such as spin qubits \cite{Samkharadze18, Landig18, Mi18}, topological materials \cite{Kroll18, Fornieri19,Pita-vidal19}, magnons \cite{Tabuchi14} or molecular electronics \cite{Bogani08,Bienfait16_NN,Bienfait2016_N,Godfrin17}, benefit from hybrid architectures where superconducting circuits can provide unique functionalities, in particular dispersive readout \citep{Blais04} and high impedance couplers. The success of superconducting circuits is linked to the availability of non-linear elements with high intrinsic coherence, namely Josephson junctions (JJs) fabricated by thermal oxidation from thin film aluminum (Al) \cite{Paik11}. However, their applicability in hybrid systems \cite{Kubo11, Ranjan13} is limited by the low critical field of Al \cite{Meservey71} and by the emergence of quantum interference effects in the JJs, even for magnetic fields aligned in-plane \cite{Schneider19}. Here we show that the JJ can be replaced by a small volume of granular aluminum (grAl) \cite{Pracht16} providing enough non-linearity to implement a superconducting transmon qubit \cite{Koch07}, which we operate in magnetic fields up to $\sim0.1\,\mathrm{T}$.
 
Granular aluminum, similarly to other materials such as NbN \cite{Niepce19}, NbTiN \cite{Samkharadze16} or TiN \cite{Shearrow18}, is an attractive choice for superconducting hybrid systems operating at radio-frequencies, due to its large critical magnetic field \cite{Cohen68}, high coherence in the microwave domain \cite{Gruenhaupt18, Zhang19, Gruenhaupt19, Kamenov20} and intrinsic non-linearity \cite{Maleeva18,Schoen20}. The constituent Al grains, about $3-5\,\mathrm{nm}$ in diameter \cite{Deutscher73}, are separated by thin oxygen barriers, therefore grAl structures can be modeled as arrays of JJs \cite{Maleeva18}. Their kinetic inductance is tunable over orders of magnitude up to $\mathrm{nH} / \square$ \cite{Valenti19}. 

Similar to JJ arrays, the non-linearity of the grAl kinetic inductance stems from the Josephson coupling between neighboring grains, and it is inversely proportional to the critical current density $j_\mathrm{c}$ and the volume of the film $V_\mathrm{grAl}$\cite{Maleeva18}. This non-linearity gives rise to a frequency shift $K$ of the fundamental plasmon mode $\omega_1$ for each added photon $(n \rightarrow n+1)$. Although the values $K(n)$ depend on the transition number $n$, to lowest order they can be approximated by a constant self-Kerr coefficient $K = \mathcal{C} \pi e a \frac{\omega_1^2}{j_\mathrm{c} V_\mathrm{grAl}}$, where $\mathcal{C}$ is a numerical factor close to unity that depends on the current distribution, $e$ is the electron charge, and $a$ is the grain size \cite{Maleeva18}.  

\begin{figure}[!t]
\begin{center}
\includegraphics[width = 1\columnwidth]{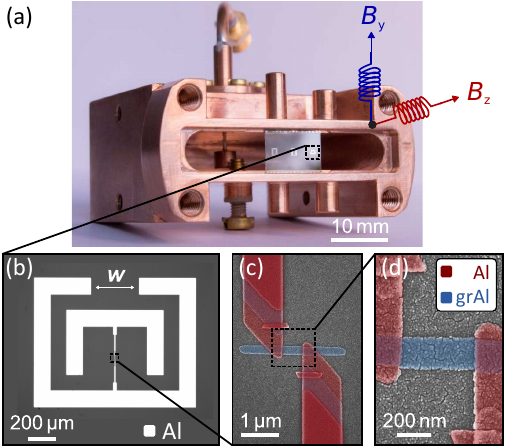}
\caption{\textbf{Sample design.} \textbf{a)} Photograph of a copper-waveguide sample holder equipped with a microwave port similar to Ref.\,\cite{Kou18}, and, optionally, with a 2D vector magnet (cf. App.\,\ref{ASEC:2D_vector_magnet}). The vector magnet is schematically represented by the blue and red coils, oriented along the $y$ and $z$ directions. The sample is positioned in the center of the waveguide and couples to its electric field along the $y$ direction. \textbf{b)} Optical image of the qubit sample, consisting of two Al pads forming a capacitor $C_\mathrm{s} \approx 137\,\mathrm{fF}$, connected by a grAl inductor $L_\mathrm{k} \approx 2.9\,\mathrm{nH}$. We adjust the coupling of the sample to the waveguide by changing the gap $w$ (cf. App.\,\ref{ASEC:FEM_Simulation}). \textbf{c) and d)} Scanning electron microscope (SEM) image of the grAl inductor (false-colored in blue) with volume $V_\mathrm{grAl} = 10 \times 200 \times 500 \,\mathrm{nm^3}$ and the Al leads (false-colored in red). The grainy surface structure is due to the antistatic Au layer used for imaging. The circuit is obtained in a single lithography step by performing a three-angle shadow evaporation. The Al layer shunts the grAl film in all areas, except for the volume $V_\mathrm{grAl}$ in the center, which constitutes the source of non-linearity for the qubit \cite{Maleeva18}. The geometric inductance is $L_\mathrm{s} = 0.45\,\mathrm{nH}$, and the contacts contribute to the $L_\mathrm{k}$ with $0.13\,\mathrm{nH}$ (cf. App.\,\ref{ASEC:Magnetic_field_dependence}). }
\label{fig_sample}
\end{center}
\end{figure}

By reducing the grAl volume and the critical current density, one can potentially increase $K(1)$ to a value much larger than the transition linewidth $\Gamma_{01}$, allowing to map a qubit to the first two levels $\ket{0}$ and $\ket{1}$, similar to a transmon qubit \cite{Koch07}. Following this approach, we construct a circuit with a transition frequency $f_1 = 7.4887\,\mathrm{GHz}$ by connecting an Al capacitor to a small volume of grAl, $V_\mathrm{grAl} = 10 \times 200 \times 500 \,\mathrm{nm^3}$, with critical current density $j_\mathrm{c} \approx 0.4\,\mathrm{mA}/\si{\micro\metre^2}$ (cf. Fig.\,\ref{fig_sample}). For this structure the estimated anharmonicity $\alpha = K(1)$ is in the MHz range \cite{Maleeva18}. Indeed, as we show below, the measured value is $\alpha = 2\pi\times 4.48\,\mathrm{MHz}$, which is much larger than the transition linewidth $\Gamma_{01} = 2 \pi \times 50 \, \mathrm{kHz}$, effectively implementing a relatively low anharmonicity transmon qubit.

Figure\,\ref{fig_sample} shows a typical copper waveguide sample holder, together with the circuit design of our qubit, consisting of a grAl film shunted by an Al capacitor. From finite-element simulations we extract a shunt capacitance $C_\mathrm{s} \approx 137\,\mathrm{fF}$ and a geometric stray inductance $L_\mathrm{s} \approx 0.45\,\mathrm{nH}$ (cf. App.\,\ref{ASEC:FEM_Simulation}). The outer electrode of the capacitor surrounds the inner electrode almost completely, except for a gap of width $w$ (cf. Fig.\,\ref{fig_sample}b), which is used to tune the coupling rate $\kappa$ between the qubit and the waveguide sample holder (cf. App.\,\ref{ASEC:FEM_Simulation}). 

The sample is fabricated on a sapphire wafer in a single-step lithography by performing a three-angle shadow evaporation. First, a $10\,\mathrm{nm}$ thick grAl layer with room temperature resistivity $\rho_\mathrm{n} = 1800 \pm 200\,\si{\micro \ohm}\,\mathrm{cm}$ and corresponding critical temperature $T_\mathrm{c} = 1.9\,\mathrm{K}$ is deposited at zero-angle, followed by two $40\,\mathrm{nm}$ thick Al layers evaporated at $\pm 35^\circ$ (cf. App.\,\ref{ASEC:AFM}). Thanks to this procedure, only a small grAl volume, highlighted in blue in Fig.\,\ref{fig_sample}d, remains unshunted by the pure Al layers and participates in the electromagnetic mode with a kinetic inductance $L_\mathrm{K} = 2.9\,\mathrm{nH}$, constituting $87\,\%$ of the total inductance.

\begin{figure}[!t]
\begin{center}
\includegraphics[width = 1\columnwidth]{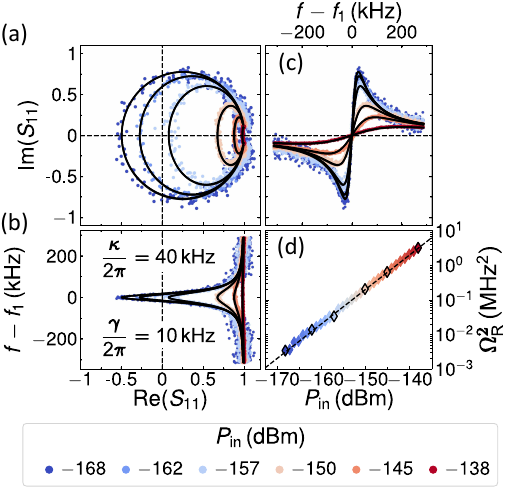}
\caption{\textbf{Resonance fluorescence.} \textbf{a)} Single-port reflection coefficient $S_{11}$ measured around the qubit frequency $f_1 = 7.4887\,\mathrm{GHz}$. For probe powers $P_\mathrm{in}$ well below the single-photon regime ($\bar{n} \ll 1$), $S_{11}$ closely resembles a circle in the quadrature plane (dark blue markers), from which, using Eq.\,\ref{EQ:reflection}, we extract the external and internal decay rates $\kappa = 2 \pi \times 40\,\mathrm{kHz}$ and $\gamma = 2 \pi \times 10\,\mathrm{kHz}$, respectively. In \textbf{b)} and \textbf{c)} we show the real and imaginary part of the reflection coefficient $\mathrm{Re}(S_{11})$ and $\mathrm{Im}(S_{11})$, respectively, as a function of the detuning between the probe frequency $f$ and the qubit frequency $f_1$. When increasing the probe power $P_\mathrm{in}$, the response becomes elliptic in the quadrature plane, which is the signature of resonance fluorescence of a two-level-system \cite{Astafiev10}. The black lines indicate fits to the experimental data according to Eq.\,\ref{EQ:reflection}. The only fitting parameter is the Rabi-frequency $\Omega_\mathrm{R}$; $\kappa$ and $\gamma$ are fixed by the fit to the low power response (cf. panel a). In \textbf{d)} we show $\Omega_\mathrm{R}^2$ as a function of incident on-chip power $P_\mathrm{in}$. For a two level system, given by the limit $\Omega_\mathrm{R} \ll \alpha$, we expect a linear dependence, as confirmed by the black dashed line passing through the coordinate origin.}
\label{fig_fluorescence}
\end{center}
\end{figure}

\begin{figure*}[!t]
\begin{center}
\includegraphics[width = 2\columnwidth]{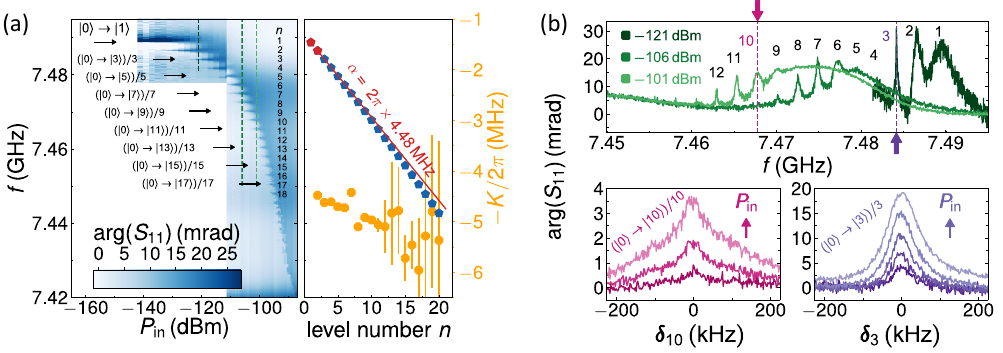}
\caption{\textbf{Energy spectrum.} \textbf{a)} Phase of the measured reflection coefficient $\arg(S_{11})$ as a function of probe frequency $f$ and incident on-chip power $P_\mathrm{in}$ (left panel). With increasing probe power we observe multi-photon transitions at frequencies $f_n$, labeled $(\ket{0} \rightarrow \ket{n}) / n$ where $n$ denotes the level number, which are almost equidistant in frequency, as plotted in the right hand panel. For clarity we only highlight with arrows the odd $n$ transitions; all indices $n$ are listed on the right hand side of the 2D plot. From the first two points in the right hand panel (highlighted in red) we extract the qubit anharmonicity $\alpha = K(1) = 2 \pi \times 4.48\,\mathrm{MHz}$, much larger than the total linewidth $\kappa + \gamma = 2\pi \times 50\,\mathrm{kHz}$ (cf. Fig.\,\ref{fig_fluorescence}). To highlight the change in $K(n)$ with increasing level number $n$ (see main text), the red line shows a linear extrapolation from the first two points. The extracted values $K(n)$ are plotted in orange using the right hand axis. \textbf{b)} Three individual measurements (top panel) performed at different probe powers ($P_\mathrm{in} = -121, -106$ and $-101\,\mathrm{dBm}$), as indicated by the vertical dashed lines in the 2D plot in a). Several multi-photon transitions are visible. With increasing power the linewidth of these transitions broadens, as shown in the bottom panels for $n = 3$ (right panel, $P_\mathrm{in} = -127$ to $-123\,\mathrm{dBm}$) and $n = 10$ (left panel, $P_\mathrm{in} = -105.75$ to $-104.75\,\mathrm{dBm}$). Here, $\delta_n = f - f_n$, with $f_3 = 7.4842\,\mathrm{GHz}$ and $f_{10} = 7.4678\,\mathrm{GHz}$, as indicated by the arrows in the top panel. In App.\,\ref{ASEC:Kerr_hamiltonian} we show that these experimental results can be quantitatively reproduced by a master-equation simulation.}
\label{fig_multi_photon}
\end{center}
\end{figure*}

We characterize the grAl transmon by performing a single-port measurement of the complex reflection coefficient $S_{11}$ as a function of probe frequency $f$, in the vicinity of the resonant frequency $f_1$ (cf. Fig.\,\ref{fig_fluorescence}). In the limit of weak driving, $\Omega_\mathrm{R} \ll \alpha$, where $\Omega_\mathrm{R}$ is the Rabi frequency, we can treat the transmon as a two-level system. If the decoherence rate is dominated by the energy relaxation rate $\Gamma_{01}$, similarly to Ref.\,\citep{Astafiev10} the complex reflection coefficient is
\begin{equation}
S_{11} (\Delta) = 1 - \frac{2 \kappa}{\Gamma_{01}} \frac{1 + i 2 \Delta / \Gamma_{01}}{1 + (2 \Delta / \Gamma_{01})^2 + 2 (\Omega_\mathrm{R} / \Gamma_{01})^2},
\label{EQ:reflection}
\end{equation} 
where, $\Delta = \omega_0 - \omega$ is the frequency detuning between the drive and the qubit frequency, and $i$ is the unit imaginary number. In contrast to a harmonic oscillator, the reflection coefficient of a qubit deviates from a circle in the quadrature plane and becomes increasingly elliptic with drive power (cf. App.\,\ref{ASEC:Resonance_fluorescence}). 

Figure\,\ref{fig_fluorescence} depicts the measured reflection coefficient $S_{11}$ as a function of qubit-drive detuning for incident on-chip powers $P_\mathrm{in}$ ranging between $-168\,\mathrm{dBm}$ and $-138\,\mathrm{dBm}$. From a least-square fit to Eq.\,\ref{EQ:reflection} (solid black lines), we extract the qubit frequency $f_1 = 7.4887\,\mathrm{GHz}$, as well as the internal loss and external coupling rates $\gamma = 2 \pi \times 10\,\mathrm{kHz}$ and $\kappa = 2 \pi \times 40\,\mathrm{kHz}$, respectively. The corresponding energy relaxation times due to radiation into the waveguide \cite{Purcell46} and internal losses are $T_\mathrm{1,\kappa} \approx 4\,\si{\micro s}$ and $T_\mathrm{1,\gamma} \approx 16\,\si{\micro s}$, respectively. As shown in Fig.\,\ref{fig_fluorescence}d, the Rabi frequency shows a linear dependence with drive amplitude, as expected for a two-level system. From a linear fit passing through the coordinate origin \cite{Astafiev10}, we calibrate the attenuation of the input line to $103\,\mathrm{dB}$, within $3\,\mathrm{dB}$ from room-temperature estimates. 

For drive powers $P_\mathrm{in} > - 138\,\mathrm{dBm}$, we observe additional features in the reflection coefficient $S_{11}$ emerging at frequencies $f_n$ below the qubit frequency $f_{1}$ (cf. Fig.\,\ref{fig_multi_photon}a left hand panel). Similar to the high power spectroscopy of JJ transmon qubits \cite{Schuster07,Bra15}, these features are multi-photon transitions into higher energy eigenstates $E_n$ starting from the ground state $E_0$, observed at frequencies $f_n = (E_n - E_0) / (n h)$, where $n$ is the level number (see App.\,\ref{ASEC:Kerr_hamiltonian} for numerical simulations of the spectrum, and App.\,\ref{ASEC:Two_tone} for two-tone spectroscopy). From the frequency detuning between the first two transitions (red markers in Fig.\,\ref{fig_multi_photon}a right hand panel), we extract a qubit anharmonicity $\alpha = 2 \pi \times 4.48\,\mathrm{MHz}$. 

Generally, for a JJ transmon the anharmonicity is given by the charging energy $E_\mathrm{c,s} = e^2 / 2 C_\mathrm{s}$ associated with the shunt capacitance \cite{Koch07}, which for our geometry is $E_\mathrm{c,s} / \hbar = 2 \pi \times 141\,\mathrm{MHz}$. In the case of an array of $N$ JJs, the anharmonicity is reduced by $N^2$ \cite{Eichler14, Siv19}, which implies $N = \sqrt{E_\mathrm{c,s} / \hbar \alpha}\approx 6$ for the JJ array implemented by our grAl volume \cite{Maleeva18}. The corresponding effective junctions are therefore separated by $\sim80\,\mathrm{nm}$, spanning approximately ten grains. This result is in agreement with recent scanning tunneling microscopy measurements performed on similar grAl films, which evidenced the collective charging of clusters of grains \cite{Yang19}.

\begin{figure*}[!t]
\begin{center}
\includegraphics[width = 2\columnwidth]{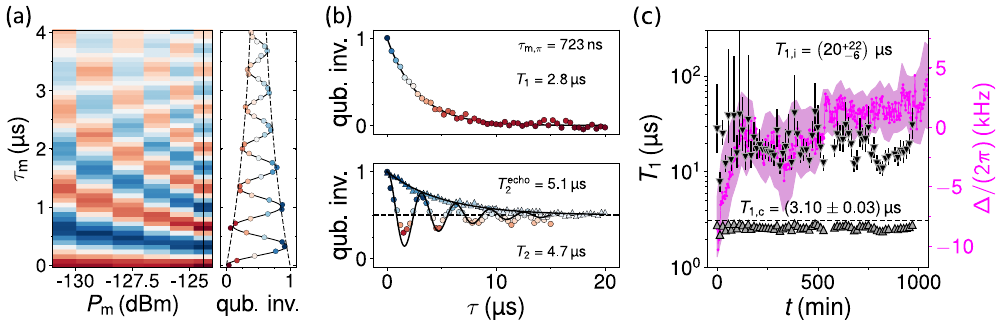}
\caption{\textbf{Time-domain manipulation and measurements of the grAl transmon. a)} Rabi-oscillations between the qubit's ground $\ket{0}$ and first excited state $\ket{1}$ as a function of the average drive power of the Gaussian shaped manipulation pulse $P_\mathrm{m}$, and its duration $\tau_\mathrm{m}$. The color scale represents the qubit population inversion from equilibrium (red) to fully inverted (blue). With increasing drive power, the Rabi-frequency increases in agreement with the spectroscopy measurement (cf. Fig.\,\ref{fig_fluorescence} and App.\,\ref{ASEC:Resonance_fluorescence}). The right-hand panel shows the Rabi-oscillation for $P_\mathrm{m} = -124.7\,\mathrm{dBm}$, highlighted in the 2D plot by a black arrow. The black dashed lines indicate an exponentially decaying envelop. \textbf{b)} Measurement of the grAl transmon coherence times. The duration of the $\pi$-pulse is $723\,\mathrm{ns}$ at a manipulation power $P_\mathrm{m} = -131.4\,\mathrm{dBm}$. The markers in the top panel show the measured population inversion at a time $\tau$ after the $\pi$-pulse, and the black line indicates an exponential fit with a characteristic energy relaxation time $T_1 = 2.8\,\si{\micro\second}$. The bottom panel shows the results of a Ramsey-fringes (circles) and an Hahn-echo (triangles) measurement. The observed frequency of the Ramsey fringes agrees within $5\%$ with the frequency detuning $\Delta_\mathrm{m} = 2 \pi \times 300\,\mathrm{kHz}$ of the $\pi$-pulse. From the fits indicated by solid lines, we extract coherence times $T_2 = 4.7\,\si{\micro\second}$ and $T_2^\mathrm{echo} = 5.1\,\si{\micro\second}$, as well as the corresponding pure dephasing times $28\,\si{\micro\second}$ and $46\,\si{\micro\second}$, respectively. \textbf{c)} Time stability of the energy relaxation time (grey triangles) and the deviation $\Delta$ of the qubit transition frequency from its average (pink markers corresponding to the right-hand axis). The pink shaded area is the uncertainty for the qubit frequency measurement. The inverted black triangles show the intrinsic energy relaxation time $T_{1,\gamma} = T_1 T_{1,\kappa}/(T_{1,\kappa} - T_1)$, where $T_{1,\kappa} = 3.1\,\si{\micro\second}$ (dashed line) is the limit due to spontaneous emission into the waveguide. Notably, the $T_1$ and frequency stability data were taken in different cool downs. }
\label{fig_time_domain}
\end{center}
\end{figure*}

From the measured multi-photon transition frequencies $f_n$, we calculate the non-linear frequency shift ${\hbar K(n) =  (E_n - E_{n-1}) - (E_{n+1} - E_n)}$ with $E_n / h= n f_n$. As shown in the right hand panel of Fig.\,\ref{fig_multi_photon}a (right hand axis), we find that $K(n)$ monotonically increases with $n$, likely due to the contribution of higher order terms in the expansion of the Josephson potential, currently not included in the model\cite{Maleeva18}.  

In the top panel of Fig.\,\ref{fig_multi_photon}b we show measurements for three different drive powers $P_\mathrm{in} = -121, -106$ and $-101\,\mathrm{dBm}$. At any drive power in this range several multi-photon peaks are visible. The linewidth of each transition broadens with power, as illustrated in the bottom panel of Fig.\,\ref{fig_multi_photon}b for the 3rd and the 10th multi-photon transition. The visibility of the peaks and the background response of the phase is in remarkable agreement with the master-equation simulation presented in App.\,\ref{ASEC:Kerr_hamiltonian}. The broadening of the $n = 10$ transition, compared to $n = 3$, can be explained by offset charge dispersion, which increases exponentially with $n$ \cite{Koch07}.

The time evolution of the qubit can also be measured using resonance fluorescence \cite{Abdumaliko11}. To enhance the signal-to-noise ratio, we added a dimer Josephson junction array amplifier (DJJAA) \cite{Winkel20}, operated in reflection, with a power gain $G_0 \gtrsim 20\,\mathrm{dB}$. In Fig.\,\ref{fig_time_domain}a we show the measured Rabi-oscillations between the qubit's ground and first excited state, obtained by applying a Gaussian shaped manipulation pulse of duration $\tau_\mathrm{m}$. For the readout, we demodulate the second half of a $3.2\,\si{\micro\second}$ long pulse with rectangular envelope and power $P_\mathrm{r} \approx -152\,\mathrm{dBm}$. Both readout and manipulation pulses are applied on resonance with the qubit transition frequency. As expected, the Rabi-frequency increases linearly with the drive amplitude.

The energy relaxation time $T_1 = 1 / \Gamma_{01}$, shown in the top panel of Fig.\,\ref{fig_time_domain}b, is measured by preparing the qubit in the excited state using a $723\,\mathrm{ns}$ $\pi$-pulse, and monitoring the dependence of the qubit population inversion as a function of the wait time $\tau$. The results are averaged over $3\times10^5$ repetitions during the course of $8\,\mathrm{min}$. The observed energy decay is fitted by a single exponential function with characteristic decay time $T_1 = 2.8\,\si{\micro\second}$. The bottom panel of Fig.\,\ref{fig_time_domain}b shows Ramsey-fringes and Hahn-echo coherence measurements, performed with the same $\pi$-pulse, yielding  $T_2 = 4.7\,\si{\micro\second}$ and $T_2^\mathrm{echo} = 5.1\,\si{\micro\second}$, respectively. The corresponding pure dephasing times are $28\,\si{\micro\second}$ and $46\,\si{\micro\second}$. These measurements reveal energy relaxation as the main decoherence mechanism, justifying the limit $T_2 \approx 2 T_1$ used for the derivation of the reflection coefficient in Eq.\,\ref{EQ:reflection}.

In Fig.\,\ref{fig_fluorescence}c, we show the fluctuations of $T_1$ and the qubit frequency monitored over $15\,\mathrm{h}$. Although spontaneous emission into the waveguide limits the energy relaxation time to $T_{1,\kappa} = 3.1\,\si{\micro\second}$, we can infer the intrinsic energy relaxation $T_{1,\gamma}$ (inverted triangles) from the measured $T_1$ values using ${T_{1,\gamma} = T_1 T_{1,\kappa}/(T_{1,\kappa} - T_1)}$. The obtained average intrinsic energy relaxation time is ${T_{1,\gamma} = 20\substack{+ 22 \\ - 6}\,\si{\micro\second}}$, consistent with the $\gamma^{-1}$ values extracted from spectroscopy (see Fig.\,\ref{fig_fluorescence}). The pink markers indicate the detuning $\Delta$ of the qubit frequency from its average value. The observed total change is on the order of the intrinsic linewidth $\gamma$. 

\begin{figure}[!t]
\begin{center}
\includegraphics[width = 1\columnwidth]{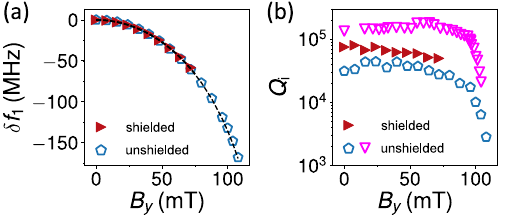}
\caption{\textbf{Magnetic field dependence. a)} Relative change in qubit frequency $\delta f_{1} = f_1(B_y) - f_{1}$ as a function of the applied in-plane magnetic field $B_y$ (see App.\,\ref{ASEC:magnetic_field_alignment} for field alignment). The experimental data was measured in two separate cooldowns: with (filled triangles) and without (open pentagons) an outer superconducting Al shield (cf. App.\,\ref{ASEC:2D_vector_magnet}). The field dependence can be fitted to a two-fluid model  \cite{Tin04} (cf. App.\,\ref{ASEC:Magnetic_field_dependence}), indicated by the black dashed line, from which we extract the thin film Al critical flux density $B_\mathrm{c,Al} = 150\pm 5 \,\mathrm{mT} $, in agreement with Ref.\,\cite{Meservey71}. A more detailed analysis of the qubit transition frequency with much higher resolution is presented in App.\,\ref{ASEC:TLS}. \textbf{b)} Internal quality factor $Q_\mathrm{i}$ for the shielded (filled triangles) and unshielded (open pentagons and triangles) case versus the in-plane magnetic field $B_y$.}
\label{fig_field_dependence}
\end{center}
\end{figure}

By applying a magnetic field $B_y$, aligned in-plane with the sample, we observe a continuous decrease of the qubit frequency $f_1 (B_{\mathrm{y}})$, as plotted in Fig.\,\ref{fig_field_dependence}a. The measurements are performed in two seperate cooldowns, with (filled crosses) and without (open pentagons) an outer superconducting Al shield. When employing the shield, the maximal field is limited to $\sim 70\,\mathrm{mT}$, after which the shield becomes affected by the field coils, introducing distortions in the field alignment. Due to the large critical field of grAl ($\sim 4 - 5 \, \mathrm{T}$ \cite{Cohen68}), the change in frequency is primarily due to the lowering of the Al gap $\Delta_\mathrm{Al}$, which leads to an increase of the kinetic inductance of the Al wires connecting the electrodes, accounting for $\sim 10\,\%$ of the total inductance.

Figure\,\ref{fig_field_dependence}b depicts the internal quality factor $Q_\mathrm{i}$ as a function of the in-plane magnetic field $B_\mathrm{y}$ measured with (filled triangles) and without (open pentagons and triangles) an outer Al shield. Compared to the data depicted in Fig.\,\ref{fig_fluorescence}, we attribute the lower internal quality factor in these three cooldowns to the removal of the $\mu$-metal shield, which likely results in an increase of the (stray) $B_z$ field. In the current design the Al pads are the most field susceptible components, rendering the qubit frequency and internal quality factor particularly sensitive to out-of-plane magnetic fields (cf. App.\,\ref{ASEC:magnetic_field_alignment}), as illustrated by the factor of three difference in $Q_\mathrm{i}$ values in the unshielded case for nominally identical setups. Finally, it is important to note that the absolute value of the non-linear frequency shift $K$ is not expected to change in magnetic field, because the ratio $\omega_1^2 / j_\mathrm{c}$ in the expression for $K$ is independent of the grAl superconducting gap. Indeed, we measure a constant $K$ up to $\sim 100\,\mathrm{mT}$ (cf. App.\,\ref{ASEC:Magnetic_field_dependence}), confirming the grAl transmon's resilience to moderate magnetic fields.

In summary, we have shown that using small volumes of grAl can provide enough non-linearity to implement magnetic field resilient superconducting qubits. We have implemented  a transmon qubit\cite{Koch07} with an anharmonicity $\alpha = 2 \pi \times 4.48\,\mathrm{MHz}$, which, although smaller than the typical values for JJ based transmons, is much larger than the qubit linewidth $\Gamma_{01} = 2 \pi \times 50\,\mathrm{kHz}$. This enables time-domain manipulation and measurement of the qubit, from which we extracted an intrinsic ${T_{1,\gamma} = 20\substack{+ 22 \\ - 6}\,\si{\micro\second}}$, and a pure dephasing time exceeding tens of microseconds. We observe multi-photon transitions to the 20th order, showcasing the robustness of the grAl transmon to external drives; a valuable asset for strongly driven quantum circuits \cite{Lescanne19}, in particular in the context of bosonic codes for quantum information \cite{Puri17,Gao19,campagneibarcq19}. Measuring the same qubit with less shielding, in the presence of in-plane magnetic fields up to $\sim70\,\mathrm{mT}$, the intrinsic linewidth remains below $\gamma = 2\pi \times 150\,\mathrm{kHz}$, limited by the pure Al capacitor pads. 

Following this proof of principle demonstration, future developments will focus on replacing all pure Al components with more field resilient materials, such as low resistivity grAl \cite{Valenti19} or Nb compounds \cite{Kwon18,Niepce19,Samkharadze16}. This will allow to operate coherent superconducting qubits in magnetic fields beyond $1\,\mathrm{T}$. In the current design, the Al pads play the role of phonon and quasiparticle traps \cite{Valenti19,Riwar19}, enhancing the coherence. In future designs, in which Al is substituted by other superconductors with higher gap and critical field, quasiparticle poisoning could be mitigated by adding dedicated phonon traps disconnected from the qubit \cite{Henriques19,Karatsu19}. In order to avoid lowering $T_1$ by the Purcell effect, following the recent non-perturbative design reported in Ref.\,\cite{Dassonneville20}, a cross-Kerr interaction to an ancilla mode could be used for readout. The limitations on qubit manipulation fidelity can be mitigated with DRAG pulses \cite{Motzoi09} and by adapting the design to decrease the grAl volume and critical current density, thereby increasing the anharmonicity. Furthermore, larger values of anharmonicity might be achieved in fluxonium qubits \cite{Manucharyan09, Gruenhaupt19}.  \\
\\
We are grateful to H. Rotzinger, U. Vool and W. Wulfhekel for fruitful discussions, and we acknowledge technical support from S. Diewald, A. Lukashenko and L. Radtke. Funding was provided by the Alexander von Humboldt foundation in the framework of a Sofja Kovalevskaja award endowed by the German Federal Ministry of Education and Research,  and by the Initiative and Networking Fund of the Helmholtz Association, within the Helmholtz Future Project \textit{Scalable solid state quantum computing.} PW and WW acknowledge support from the European Research Council advanced grant MoQuOS (N. 741276). AVU acknowledges partial support from the Ministry of Education and Science of the Russian Federation in the framework of the Increase Competitiveness Program of the National University of Science and Technology MISIS (Contract No. K2-2018-015). Facilities use was supported by the  KIT  Nanostructure  Service  Laboratory  (NSL).  We acknowledge qKit \cite{qkit} for providing a convenient measurement software framework.

\bibliography{grAl_transmon_references}



\appendix

\renewcommand{\appendixname}{Appendix}


\section{2D vector magnet}
\label{ASEC:2D_vector_magnet}
\begin{figure}[!t]
\begin{center}
\includegraphics[width = \columnwidth]{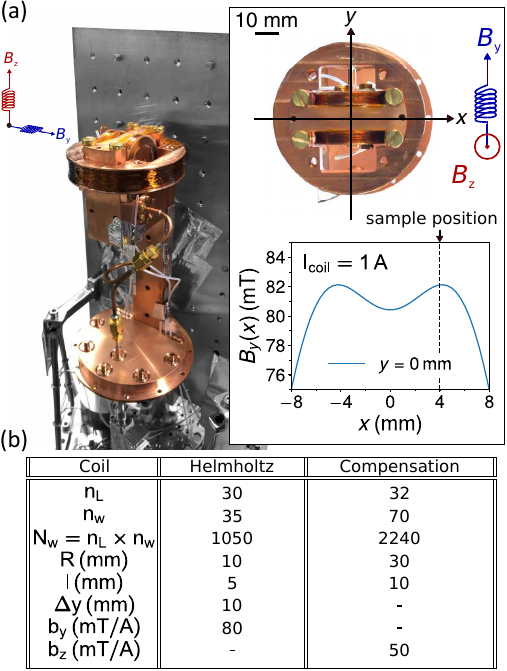}
\caption{\textbf{Photograph of the cryogenic setup used for the magnetic field measurements in Fig.\,\ref{fig_field_dependence}.} The copper waveguide sample holder equipped with the 2D vector magnet (highlighted in color) is mounted at the dilution stage of a table-top \textit{Sionludi} dilution refrigerator \cite{Sionludi12} (greyscaled), with a base temperature $T_\mathrm{base} \approx 20 - 30\,\mathrm{mK}$. The flat copper cylinder visible in the lower part of the image is the lid of the outer shield (not shown), which consists of successive copper (Cu) and aluminum (Al) cylinders, similar to Ref.\,\cite{Gruenhaupt17}. The Cu shield was used in both measurements presented in Fig.\,\ref{fig_field_dependence}, while the Al shield was only used during the "shielded" cooldown. The inset shows the top view of the copper waveguide sample holder including the 2D vector magnet (top), and the numerically calculated Helmholtz field (bottom) $B_y$ for a bias current $I_\mathrm{coil} = 1\,\mathrm{A}$ as a function of lateral position $x$. The magnetic field of the two Helmholtz coils is aligned within machining precision with the in-plane direction $y$ of the thin-films. The $B_z$ coil is the compensation coil we use to align in-situ the in-plane field. \textbf{b)} Table summarizing the geometric parameters for the coils of the 2D vector magnet: number of winding layers $n_\mathrm{L}$, number of windings per layer $n_\mathrm{w}$, total number of windings $N_\mathrm{w}$, inner coil radius $R$, coil length $l$, vertical distance between Helmholtz coils $\Delta y$, magnetic flux density in $y$ and $z$ direction per $1\,\mathrm{A}$ of bias current, $b_y$ and $b_z$, respectively.}
\label{supfig_sample}
\end{center}
\end{figure}

The waveguide sample holder can be equipped with a 2D vector magnet, which consists of a pair of Helmholtz coils (HH) and a solenoid for field alignment, denoted compensation coil in the following (cf. Fig.\,\ref{supfig_sample}a). The field direction of the HH coils is aligned within machining precision with the in-plane direction of the chip, $y$. The compensation field is oriented perpendicular to it, in the $z$ direction, out-of-plane with respect to the sample. All coils are winded with the same type of NbTi, multifilament superconducting wire with diameter $d = 140\,\si{\micro m}$ (Supercon. Inc. 54S43). The winding parameters - the number of layers $n_\mathrm{L}$ and the number of windings per layer $n_\mathrm{w}$ - and the physical dimensions of the coils - radius $R$, length $l$ and vertical distance $\Delta y$ (in the case of the HH coils) - are summarized in Fig.\,\ref{supfig_sample}b. 

From both coil geometries, we calculate the relation between the applied bias current $I_\mathrm{coil}$ and the magnetic flux density $\vec{B}$ at position $\vec{r}$ using the Biot-Savart law. For simplicity, we approximate the coils with $n_\mathrm{L} \times n_\mathrm{w}$ single loops. The loop radius depends on the layer number and the position along the coils symmetry axis depends on the winding number, both gradually increasing by the wire diameter $d = 140\,\si{\micro \metre}$. Following the approach of Caparelli \textit{et al.} \cite{Caparelli01}, we approximate the magnetic field components $B_r(\vec{r})$, in radial direction (parallel to the loop plane $xz$), and $B_y(\vec{r})$ by truncating the sum after 20 terms.  

Figure\,\ref{supfig_sample}a (bottom right) depicts the numerically calculated magnetic flux density of the HH coils as a function of the lateral position $x$ for a bias current $I_\mathrm{coil} = 1\,\mathrm{A}$ and $y = 0\,\mathrm{mm}$. The center of the waveguide is the origin for $x$ and $y$ as indicated by the black dashed lines (top right panel). Since the magnetic flux density $B_y$ changes by only $3\%$ in the region where the sample chip is mounted ($-5\,\mathrm{mm} \leq x \leq 5\,\mathrm{mm}$), we take the value $b_y = 80\,\mathrm{mT/A}$ to convert the HH bias current $I_\mathrm{coil}$ into a magnetic flux density. For the compensation coil we find a conversion factor $b_z = 50\,\mathrm{mT/A}$.

\begin{figure*}[t]
\begin{center}
\includegraphics[width = 2\columnwidth]{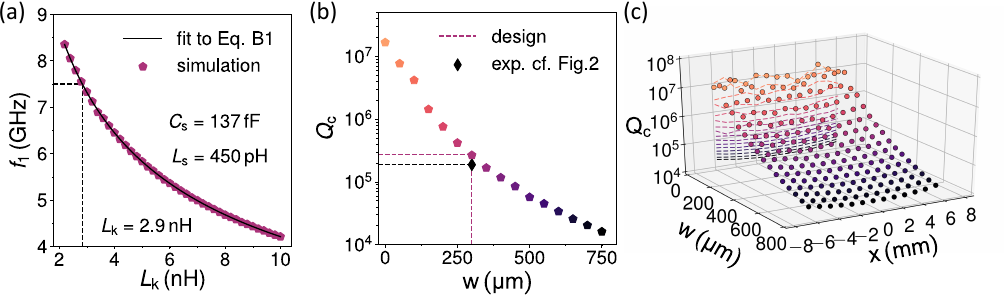}
\caption{\textbf{Finite-element method simulations of the linearized circuit.} \textbf{a)} Simulated transition frequency $f_1$ of the linearized circuit obtained using the eigenmode solver of a commercial finite-element method simulator (HFSS). For the simulation, the grAl volume is substituted with a lumped-element inductance $L_\mathrm{k}$ which we sweep from $2\,\mathrm{nH}$ to $10\,\mathrm{nH}$ in order to fit the shunt capacitance $C_\mathrm{s}$ and additional geometric inductance $L_\mathrm{s}$ of our design using Eq.\,\ref{EQ:f0simulation}. We extract $C_\mathrm{s} = 137\,\mathrm{fF}$ and $L_\mathrm{s} = 450\,\mathrm{pH}$. As indicated by the dotted cursors, the measured frequency $f_1 = 7.4887\,\mathrm{GHz}$ corresponds to a value $L_\mathrm{k} = 2.9\,\mathrm{nH}$. \textbf{b)} Simulated external quality factor $Q_\mathrm{c}$ as a function of the gap $w$ in the outer electrode for $L_\mathrm{k} = 2.9\,\mathrm{nH}$. Similarly to the design of Ref.\,\cite{Richer17}, by closing the gap we can vary the external quality factor by three orders of magnitude. For clarity (cf. panel c), the color of the markers is related to the value of $w$. The colored dashed lines indicate the design value for the experiment and the black marker indicates the measured external quality factor $Q_\mathrm{c} = 1.9 \times 10^5$ (cf. Fig.\,\ref{fig_fluorescence}). \textbf{c)} Simulated external quality factor as a function of the lateral chip position $x$ and gap width $w$. The value of $x$ is measured from the center of the waveguide to the symmetry axis of the circuit. Due to the large aspect ratio of the waveguide's cross section ($30\,\mathrm{mm} \times 6\,\mathrm{mm} $), the external quality factor varies by less than a factor of 2 along $x$.}
\label{supfig_simulation}
\end{center}
\end{figure*}

\section{Finite-element method simulations}
\label{ASEC:FEM_Simulation}
The eigenfrequency and external coupling rate of the circuit to the waveguide sample holder is simulated with a commercial finite-element method simulator (HFSS - High frequency structure simulator). The inductive contribution of the granular aluminum (grAl) volume is modeled with a linear lumped-element inductor $L_\mathrm{K}$. Capacitive contributions arising from the grAl microstructure are not considered. In order to extract the lumped-element shunt capacitance $C_\mathrm{s}$ and geometric stray inductance $L_\mathrm{s}$ of our circuit design, we sweep the inductance $L_\mathrm{K}$ while keeping the design geometry fixed (cf. Fig.\,\ref{supfig_simulation}a). The dimensions of the circuit geometry are listed in Tab.\,\ref{tab:sample_geometry}. The simulated eigenfrequencies are fitted to the lumped-element model  
\begin{equation}
f_1 (L_\mathrm{K}) = \frac{1}{2 \pi \sqrt{C_\mathrm{s} (L_\mathrm{K} + L_\mathrm{s})}},
\label{EQ:f0simulation}
\end{equation}
yielding $C_\mathrm{s} = 137\,\mathrm{fF}$ and $L_\mathrm{s} = 450\,\mathrm{pH}$ (cf. Fig.\,\ref{supfig_simulation}a). 

The external quality factor can be varied over three orders of magnitude by changing the gap $w$ in the outer electrode (cf. Fig.\,\ref{fig_sample}b main text and Fig.\,\ref{supfig_simulation}b). As shown in Fig.\,\ref{supfig_simulation}c, we confirm that $Q_\mathrm{c}$ does not significantly change with the lateral position $x$, up to $x = \pm 8 \, \mathrm{mm}$. The measured sample is shifted by $x = 4\,\mathrm{mm}$ from the center.

\begin{table}[t]
\begin{center}
\caption{\textbf{Geometrical parameters of the sample (cf. Fig.\,\ref{fig_sample}b):} sample width $b$, sample height $h$, gap in outer electrode $w$, length of the bridge connecting the capacitor pads $l_\mathrm{b}$, width of capacitor pads $w_\mathrm{f}$, gap between capacitor pads $g_\mathrm{f}$.}
\begin{tabular}{|| c c c c c c ||} 
\hline
$b$ & $h$  & $w$ & $l_\mathrm{b}$ & $w_\mathrm{f}$ & $g_\mathrm{f}$\\ 
$\si{(\micro \metre)}$ & $\si{(\micro \metre)}$ & $\si{(\micro \metre)}$ & $\si{(\micro \metre)}$ & $\si{(\micro \metre)}$ & $\si{(\micro \metre)}$ \\ \hline \hline 1000 & 800 & 300 & 400 & 100 & 100 \\ \hline
\end{tabular}
\end{center}
\label{tab:sample_geometry}
\end{table}

\section{Resonance fluorescence: reflection coefficient}
\label{ASEC:Resonance_fluorescence}
In the limit of weak driving $\Omega_\mathrm{R} \ll \alpha$, we describe our transmon \cite{Koch07} as an effective two-level system \cite{Astafiev10}. Under this assumption, the reflection coefficient is 
\begin{equation}
S_{11} = 1 - \sqrt{\kappa} \frac{\braket{\sigma_-}}{\alpha_\mathrm{in}},
\label{AEQ:reflection_coefficient}
\end{equation}
where $\kappa$ is the single-photon coupling rate to the waveguide, $\braket{\sigma_-}$ is the expectation value of the lowering operator in the two dimensional qubit subspace $\{\ket{0} , \ket{1}\}$, and $\alpha_\mathrm{in}$ is the expectation value of the annihilation operator of the incident bosonic single-mode field amplitude $a_\mathrm{in}$ 
\cite{Cottet19}
. Here, we assume a classical drive $a_\mathrm{in} = \alpha_\mathrm{in} e^{-i\omega t}$, which has in general a complex amplitude $\alpha_\mathrm{in}$.

The expectation value of the lowering operator is expressed using the Pauli operators $\sigma_x$ and $\sigma_y$: $\braket{\sigma_-} = (\braket{\sigma_x} - i \braket{\sigma_y}) / 2$. The steady state expectation values ($\mathrm{time}\,t \rightarrow \infty)$ for the Pauli operators $\sigma_x$, $\sigma_y$, $\sigma_z$ under a continuous drive with amplitude $\Omega_\mathrm{R}$ and detuning $\Delta = \omega_\mathrm{q} - \omega$, in the presence of qubit energy relaxation at rate $\Gamma_{01}$ and qubit dephasing at rate $\Gamma_2 = \Gamma_{01}/2 + \Gamma_\varphi$, ($\Gamma_\varphi:$ pure dephasing rate) are the following \cite{Cottet19}:
\begin{align}
\braket{\sigma_x (\Omega_\mathrm{R} , \Delta)} &= \Gamma_{01} \Gamma_2 \Omega_\mathrm{R} \left( \Gamma_{01} (\Gamma_2^2 + \Delta^2) + \Gamma_2 \Omega_\mathrm{R}^2  \right)^{-1} \label{AEQ:Pauli_sigmax}\\
\braket{\sigma_y (\Omega_\mathrm{R} , \Delta)} &= \Gamma_{01} \Delta \Omega_\mathrm{R} \left( \Gamma_{01} (\Gamma_2^2 + \Delta^2) + \Gamma_2 \Omega_\mathrm{R}^2  \right)^{-1} \label{AEQ:Pauli_sigmay}\\
\braket{\sigma_z (\Omega_\mathrm{R} , \Delta)} &= - 1 +  \Gamma_2 \Omega_\mathrm{R}^2 \left( \Gamma_{01} (\Gamma_2^2 + \Delta^2) + \Gamma_2 \Omega_\mathrm{R}^2  \right)^{-1}
\label{AEQ:Pauli_sigmaz}
\end{align}
Using Eq.\,\ref{AEQ:Pauli_sigmax} and Eq.\,\ref{AEQ:Pauli_sigmay},
\begin{equation}
\braket{\sigma_-} = \frac{1}{2} \frac{\Gamma_{01} \Gamma_2 \Omega_\mathrm{R} - i \Gamma_{01} \Delta \Omega_\mathrm{R}}{\Gamma_{01} \left( \Gamma_2^2 + \Delta^2 \right) + \Gamma_2 \Omega_\mathrm{R}^2}.
\label{AEQ:Pauli_sigma_minus_explizit}
\end{equation}
Inserting Eq.\,\ref{AEQ:Pauli_sigma_minus_explizit} into Eq.\,\ref{AEQ:reflection_coefficient}, the reflection coefficient writes
\begin{equation}
S_{11} (\Delta) = 1 - \frac{\sqrt{\kappa} \Omega_\mathrm{R}}{2 \alpha_\mathrm{in}} \frac{\Gamma_{01} \Gamma_2 - i \Gamma_{01} \Delta}{\Gamma_{01} (\Gamma_2^2  +\Delta^2) + \Gamma_2 \Omega_\mathrm{R}}. 
\label{AEQ:reflection_coefficient_complete}
\end{equation}
The relation between the Rabi frequency $\Omega_\mathrm{R}$ and the drive amplitude $\alpha_\mathrm{in}$ is 
\begin{equation}
\Omega_\mathrm{R} = 2 \sqrt{\kappa} \braket{a_\mathrm{in}} = 2 \sqrt{\kappa} \alpha_\mathrm{in}.
\label{AEQ:Rabi_frequency}
\end{equation}
Using Eq.\,\ref{AEQ:Rabi_frequency} and in the limit of negligible pure dephasing $\Gamma_\varphi \ll \Gamma_{01}$, Eq.\,\ref{AEQ:reflection_coefficient_complete} simplifies to 
\begin{equation}
S_{11} (\Delta) = 1 - \frac{2 \kappa}{\Gamma_{01}} \frac{1 + i 2 \Delta / \Gamma_{01}}{1 + (2 \Delta / \Gamma_{01})^2 + 2 (\Omega_\mathrm{R} / \Gamma_{01})^2},
\label{AEQ:reflection_coefficient_Purcell}
\end{equation} 
which is Eq.\,\ref{EQ:reflection} in the main text. The factor in front of the second term of Eq.\,\ref{AEQ:reflection_coefficient_Purcell} is the coupling efficiency $\kappa / \Gamma_{01}$, with $\Gamma_{01} = \kappa + \gamma$. It is a measure of the relative size of the internal loss rate $\gamma$ compared to the external coupling rate $\kappa$. 

\begin{figure}[t!]
\begin{center}
\includegraphics[width = 1\columnwidth]{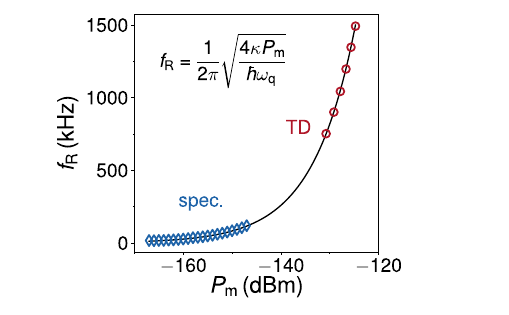}
\caption{\textbf{Rabi frequency $f_\mathrm{R} = \Omega_\mathrm{R} / (2 \pi)$ from spectroscopy and time-domain measurements.} The blue markers (labeled spec.) indicate the Rabi frequency extracted from fits to the frequency dependence of the reflection coefficient according to Eq.\,\ref{AEQ:reflection_coefficient_Purcell}, similar to Fig.\,\ref{fig_fluorescence}, while the red markers indicate the values extracted from the time-domain (TD) measurements shown in Fig.\,\ref{fig_time_domain}a. Both spectroscopy and time-domain data sets were taken in the same cool down (run \#7). The black solid line represents the expected Rabi frequency for an ideal qubit according to Eq.\,\ref{AEQ:Rabi_frequency}, provided drive power is $P_\mathrm{in} = P_\mathrm{m} = \hbar \omega_q |\alpha_\mathrm{in}|^2$.  }
\label{supfig_Rabi}
\end{center}
\end{figure}

Figure\,\ref{supfig_Rabi} shows the Rabi frequency extracted from spectroscopy, similar to Fig.\,\ref{fig_fluorescence}, and the time-domain measurements shown in Fig.\,\ref{fig_time_domain}a. The measurements follow the dependence expected for a two level system given by Eq.\,\ref{AEQ:Rabi_frequency}.

\begin{figure}[!t]
\begin{center}
\includegraphics[width = 1\columnwidth]{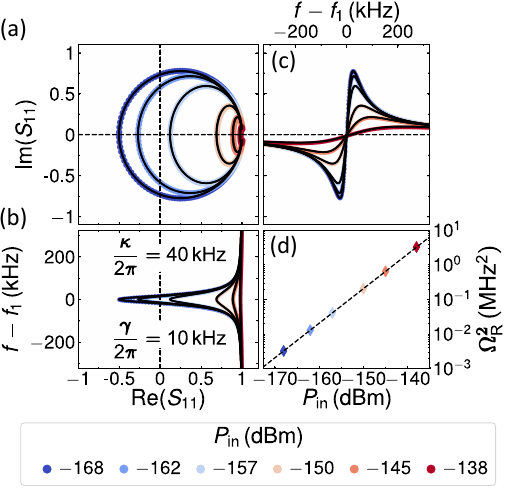}
\caption{\textbf{Numerical simulations of resonance fluorescence.} Complex reflection coefficient $S_{11}$ \textbf{(a)}, its real part \textbf{(b)} and imaginary part \textbf{(c)}, numerically calculated for probe frequencies $f$ around the resonance frequency {$f_1 = 7.4887\,\mathrm{GHz}$} of the Kerr Hamiltonian in Eq.\,\ref{AEQ:Kerr_hamiltonian}. For comparison, the input drive power $P_\mathrm{in}$, the Kerr-coefficient {$K  =2 \pi \times 4.5\,\mathrm{MHz}$}, and the decay rates {$\gamma = 2 \pi \times 10\,\mathrm{kHz}$} and {$\kappa = 2 \pi \times 40\,\mathrm{kHz}$}, are set to be the same as in the experiment. The black lines indicate least-square fits using Eq.\,\ref{EQ:reflection}. The only fit parameter is the Rabi-frequency $\Omega_\mathrm{R}$ reported in d. The linear dependence of the Rabi frequency with drive amplitude and the quantitative agreement with the measured data shown in Fig.\,\ref{fig_fluorescence}d confirm the validity of the qubit limit and the value of the input line attenuation $A = 103\,\mathrm{dB}$.}
\label{supfig_fluorescence}
\end{center}
\end{figure}

\begin{figure*}[!t]
\begin{center}
\includegraphics[width = 2\columnwidth]{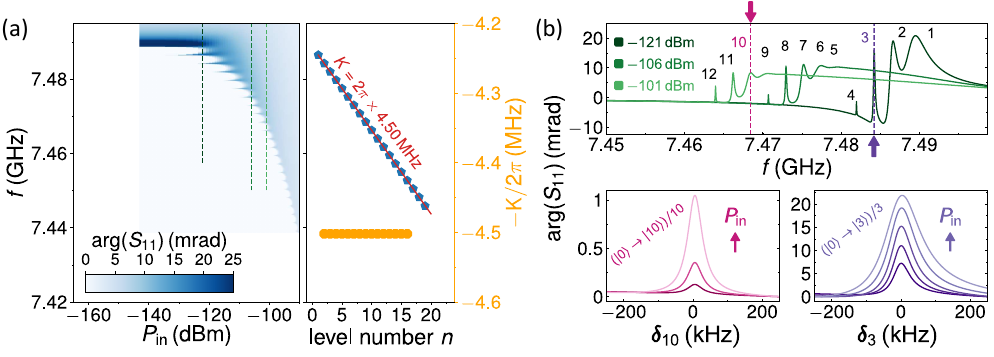}
\caption{\textbf{Numerical calculation of the energy spectrum.} \textbf{a)} Phase of the numerically calculated reflection coefficient $\arg(S_{11})$ as a function of probe frequency $f$ and probe power $P_\mathrm{in}$ (left panel) according to App.\,\ref{ASEC:Kerr_hamiltonian} and \ref{ASEC:Kerr_hamiltonian}. The multi-photon peaks are equally spaced in frequency, with the difference given by $K / 2 = 2 \pi \times 2.25\,\mathrm{MHz}$ (right panel). In contrast to the experiment, we observe a constant frequency shift $K$ (orange markers, right hand panel), as expected, because the Kerr Hamiltonian in Eq.\,\ref{AEQ:Kerr_hamiltonian} only contains terms up to the 4th order. \textbf{b)} Three individual traces (top panel) calculated at distinct drive powers ($P_\mathrm{probe} = -121, -106$ and $-101\,\mathrm{dBm}$). Similarly to the experimental results shown in Fig.\,\ref{fig_multi_photon}b, several multi-photon transitions are visible at any given drive power in this range. With increasing power, the linewidth of the transitions broadens as depicted by the zoom-ins (bottom panels) around the 3rd (right, $P_\mathrm{probe} = -127$ to $-123\,\mathrm{dBm}$) and 10th (left, $P_\mathrm{probe} = -105.75$ to $-104.75\,\mathrm{dBm}$) multi-photon transition.}
\label{supfig_multi_photon}
\end{center}
\end{figure*}

\section{Numerical calculation of the Kerr Hamiltonian}
\label{ASEC:Kerr_hamiltonian}
Since our transmon qubit has a relatively small anharmonicty $\alpha$ (cf. Fig.\,\ref{fig_multi_photon} main text), we can simulate our circuit as an anharmonic oscillator with self-Kerr coefficient $K = 2 \pi \times 4.5\,\mathrm{MHz}$. This model is further justified by the weak dependence of $K$ on the level number $n$ as measured in Fig.\,\ref{fig_multi_photon} (main text). The driven Kerr Hamiltonian expressed in the rotating frame of the coherent drive applied at frequency $\omega$ is
\begin{equation}
H_\mathrm{Kerr} / \hbar = \Delta a^\dagger a - \frac{K}{2} a^{\dagger 2} a^2 - \frac{\Omega}{2} (a^\dagger + a).
\label{AEQ:Kerr_hamiltonian}
\end{equation}
Here, $\Delta = \omega_\mathrm{1} - \omega$ is the detuning between the fundamental transition frequency $\omega_\mathrm{1}$ and the drive tone, $a^\dagger$ and $a$ are the bosonic single-mode field amplitude creation and annihilation operators, respectively, and $\Omega$ is the drive amplitude. Notably, the drive amplitude $\Omega$ corresponds to the Rabi-frequency $\Omega_\mathrm{R}$, only in the limits $K \gg \kappa$ and $\Omega \ll K$. Both criteria are met in our experiment. 

In analogy to Eq.\,\ref{AEQ:reflection_coefficient} in the case of a two-level system, the reflection coefficient of an (anharmonic) oscillator is \cite{Gardiner85, Clerk2010}
\begin{equation}
S_{11} = 1 - \sqrt{\kappa} \frac{\braket{a}}{\alpha_\mathrm{in}}.
\end{equation}
We calculate the expectation values $\braket{a}$ as a function of detuning $\Delta$ and drive amplitude $\Omega$ by solving the corresponding master equation numerically using Qutip \cite{Johannson12,Johannson13}. The master equation in Lindblad form in the presence of energy relaxation at rate $\Gamma_{01}$ is
\begin{equation}
\dot{\rho}_\mathrm{s} = - \frac{i}{\hbar} [H_\mathrm{Kerr} , \rho_\mathrm{s}] + \Gamma_1 \mathcal{D}[a](\rho_\mathrm{s}).
\end{equation}
Here, $\rho_\mathrm{s}$ is the system density matrix, $H_\mathrm{Kerr}$ is the driven Kerr Hamiltonian given in Eq.\,\ref{AEQ:Kerr_hamiltonian} and $\mathcal{D}[a](\rho)$ is the Lindblad superoperator introducing single-photon dissipation 
\begin{equation}
\mathcal{D}[a](\rho_\mathrm{s}) = a \rho_\mathrm{s} a^\dagger  - \frac{1}{2} a^\dagger a \rho_\mathrm{s} - \frac{1}{2} \rho_\mathrm{s} a^\dagger a.
\end{equation}
For our numerical calculations, we use internal and external decay rates $\gamma = 2 \pi \times 10\,\mathrm{kHz}$ and $\kappa = 2 \pi \times 40\,\mathrm{kHz}$, respectively, and a total energy relaxation rate $\Gamma_{01} = \kappa + \gamma = 2 \pi \times 50\,\mathrm{kHz}$. The dimension of the considered Hilbert space is $N_\mathrm{level} = 30$. The drive amplitudes $\Omega = 2 \sqrt{\kappa} \alpha_\mathrm{in}$ are chosen to coincide with the values used in the experiment, with $\alpha_\mathrm{in} = \sqrt{P_\mathrm{in} / (\hbar w_0)}$.  

Figure\,\ref{supfig_fluorescence} depicts the numerically calculated reflection coefficients $S_{11}$ and the corresponding least-square fits using Eq.\,\ref{EQ:reflection}. The colors are related to the drive power similarly to Fig.\,\ref{fig_fluorescence}. The linear fit to the extracted Rabi frequencies shown in Fig.\,\ref{supfig_fluorescence}d confirms that the two-level approximation (qubit limit) is valid in the parameter space of our experiment. 

We reproduce the experimental results presented in the main text in Fig.\,\ref{fig_multi_photon} by numerically calculating the reflection coefficient based on the treatment discussed in App.\,\ref{ASEC:Kerr_hamiltonian}, for the same range of probe frequency and power. Figure\,\ref{supfig_multi_photon}a depicts the phase of the reflection coefficient $\arg(S_{11})$ as a function of the incident power $P_\mathrm{in}$ and probe frequency $f$. Similar to the experiment, multi-photon transitions at frequencies $f_n = (E_n - E_0) / (n h)$ become visible. From a linear fit to the extracted multi-photon frequencies (cf. Fig.\,\ref{supfig_multi_photon}a right panel), we recover the self-Kerr coefficient $K = 2 \pi \times 4.5\,\mathrm{MHz}$ set in the calculation, as expected. Since the Kerr Hamiltonian in Eq.\,\ref{AEQ:Kerr_hamiltonian} does not contain beyond 4th order terms, in contrast with the experimental results, the simulated $K (n)$ is independent of the level number $n$. This confirms that the Hilbert-space dimension $N_\mathrm{level} = 30$ was chosen sufficiently large.  

For comparison to Fig.\,\ref{fig_multi_photon}b shown in the main text, Fig.\,\ref{supfig_multi_photon}b  shows the phase of the calculated reflection coefficient $\arg(S_{11})$ as a function of the probe frequency $f$ for three drive powers $P_\mathrm{in} = -121$, $-106$ and $-101\,\mathrm{dBm}$.

\begin{figure}[!t]
\begin{center}
\includegraphics[width = \columnwidth]{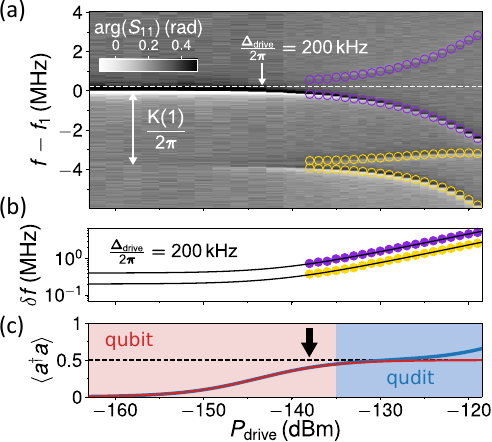}
\caption{\textbf{Two-tone spectroscopy.} \textbf{a)} Phase of the reflection coefficient $\arg(S_{11})$ measured with a weak probe tone of constant power $P_\mathrm{in} = -160\,\mathrm{dBm}$ in the vicinity of the qubit frequency $f_1$, while an additional microwave drive is applied at $\Delta_\mathrm{drive} / 2 \pi = f_\mathrm{drive} - f_1 =   200\,\mathrm{kHz}$ detuning, indicated by the dashed white line. With increasing drive power $P_\mathrm{drive}$, the fundamental transition starts to split into two distinct transitions, corresponding to the side-bands of the Mollow-triplet, at frequencies $f_\mathrm{\pm} = f_1 \pm \Omega_\mathrm{R} / 2 \pi$ \cite{Baur09,Astafiev10}. Since the population of the first excited state increases with drive power, at $P_\mathrm{drive} \geq -150\,\mathrm{dBm}$ a second transition becomes visibile at $3.9\,\mathrm{MHz}$ below $f_1$. This corresponds to the $\ket{1} \rightarrow \ket{2}$ transition. The measured anharmonicity $\alpha = 2 \pi \times 3.9\,\mathrm{MHz}$, as well as the qubit frequency $f_1 = 7.6790\,\mathrm{GHz}$ are slightly different than the values reported in the main text, due to the fact that the measurements were taken in different cooldowns. In total we performed four cooldowns, summarized in Tab.\,\ref{tab:Overview}. Similar to the fundamental transition, the second transition also splits with increasing $P_\mathrm{drive}$, an effect denoted Autler-Townes\cite{Baur09}. \textbf{b)} Extracted frequency splittings $\delta f$ for the first two transitions. The black lines correspond to the theoretical predictions $2 \Omega_\mathrm{R}$ (Mollow triplet sidebands) and $\Omega_\mathrm{R}$ (Autler-Townes), respectively, with $\Omega_\mathrm{R} = \sqrt{\Delta^2 + \Omega^2}$ and $\Omega = \sqrt{4 \kappa P_\mathrm{drive} / h f_\mathrm{drive}}$. \textbf{c)} Expectation value for the photon number operator $\braket{a^\dagger a}$ as a function of the drive power $P_\mathrm{drive}$  numerically calculated for a qubit (red line) and for an anharmonic multi-level oscillator (cf. App.\,\ref{ASEC:Kerr_hamiltonian}).}
\label{supfig_TTS}
\end{center}
\end{figure}

We would like to add that multi-photon transitions have been measured in flux qubits \cite{Oliver05,Izmalkov08} and Cooper pairs boxes \cite{Wilson07,Sillanpaa06}, with the notable difference that in these previous cases the transitions are between the ground and the first excited state \cite{Shevchenko10}, while in our case they are between the ground and higher excited states.

\section{Two-tone spectroscopy}
\label{ASEC:Two_tone}
Figure\,\ref{supfig_TTS} shows a two-tone spectroscopy of the qubit sample in the vicinity of its fundamental transition frequency $f_1$. We apply a fixed frequency drive tone at $\Delta_\mathrm{drive} = 2 \pi \times 200\,\mathrm{kHz}$ above $f_1$ with varying drive power $P_\mathrm{drive}$. Simultaneously, we measure the reflection coefficient $S_{11}$ by appling a weak probe tone at varying frequency $f$ and constant power $P_\mathrm{in} = - 160\,\mathrm{dBm}$.

For small drive powers, only a single response at the qubit frequency is visible in the phase of the reflection coefficient $\arg (S_{11})$ shown in Fig.\,\ref{supfig_TTS}a. With increasing drive power, the occupation of the first excited state $\ket{1}$ increases and a second feature becomes visible $ 3.9\,\mathrm{MHz}$ below the qubit frequency, corresponding to the single-photon transition between the first and the second excited state $\ket{1} \rightarrow \ket{2}$, and quantifying the qubit anharmonicity $\alpha$. The measured anharmonicity $\alpha = 2 \pi \times 3.9\,\mathrm{MHz}$ and the qubit frequency $f_1 = 7.6790\,\mathrm{GHz}$ are slightly different compared to the values reported in the main text, due to the fact that the measurements were taken in different cooldowns and the sample parameters changed (most likely the grAl resistivity $\rho_\mathrm{n}$). In total we performed four cooldowns, summarized in Tab.\,\ref{tab:Overview}. 

Both the $\ket{0} \rightarrow \ket{1}$ and $\ket{1} \rightarrow \ket{2}$ transitions split into two distinct transitions with increasing $P_\mathrm{drive}$. This observation is in quantitative agreement with the theoretical modelling of a driven three-level system (qutrit) \cite{Baur09,Astafiev10} (cf. Fig.\,\ref{supfig_TTS}b).

The expectation value of the photon number operator $\braket{a^\dagger a}$ is shown in Fig.\ref{supfig_TTS}c, numerically calculated for a two-level (qubit, red) and a multi-level system (qudit, blue) with constant anharmonicity (cf. App.\ref{ASEC:Kerr_hamiltonian}). In contrast to a multi-level system, for a qubit the steady state occupation number versus drive power saturates at $0.5$. The red shaded area in Fig.\,\ref{supfig_TTS}c highlights the qubit limit in which the difference between the occupation numbers of an ideal qubit and our system (main text $\alpha = 2 \pi \times 4.48\,\mathrm{MHz}$) is below $1\,\%$. The black arrow indicates the maximal drive power used for the resonance fluorescence measurements shown in Fig.\,\ref{fig_fluorescence}.

\section{Magnetic field alignment}
\label{ASEC:magnetic_field_alignment}
\begin{figure}[!t]
\begin{center}
\includegraphics[width = \columnwidth]{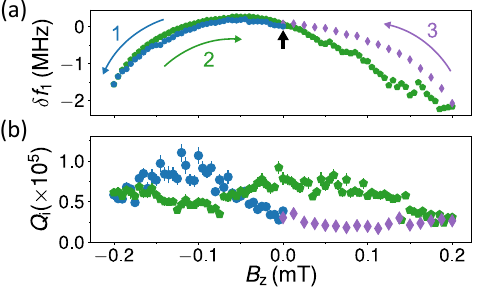}
\caption{\textbf{Qubit response to out-of-plane magnetic fields} \textbf{a)} Change in qubit frequency $\delta f_1 = f_1 (B_z) - f_{1}$ as a function of the out-of-plane compensation field $B_z$ for $B_\mathrm{y} = 0$. The colored arrows indicate the order of the measurement cycle. \textbf{b)} Internal quality factor $Q_\mathrm{i}$ as a function of $B_z$ extracted from the same measurement sequence. The internal quality factor exhibits a maximum around $B_z = -0.1\,\mathrm{mT}$, reaching a factor of 2 higher than the initial $Q_\mathrm{i}$ value in zero-field (cf. Fig.\,\ref{fig_field_dependence}b, blue pentagons)}
\label{supfig_Qis_Bz}
\end{center}
\end{figure}

Figure\,\ref{supfig_Qis_Bz}a shows the change in qubit frequency $\delta f_1 = f_1 (B_z) - f_{1}$, as a function of $B_z$ for $B_y = 0$. The field sweeps are the following: i) $0\,\mathrm{mT} \rightarrow -0.2\,\mathrm{mT}$ (blue), ii) $-0.2\,\mathrm{mT} \rightarrow 0.2\,\mathrm{mT}$ (green), iii) $0.2\,\mathrm{mT} \rightarrow 0\,\mathrm{mT}$ (purple), with the black arrow indicating the starting point. The qubit frequency decreases with increasing field magnitude and returns to its initial value when the field is swept in the opposite direction. For fields applied in positive $z$-direction, we observe several jumps and a much less smooth change in the qubit frequency with field. This observation is not strictly related to the positive $z$-direction, but depends on the order of the measurement sequence. Figure\,\ref{supfig_Qis_Bz}b shows the internal quality factor $Q_\mathrm{i}$ extracted from the same measurement sequence as the qubit frequency shown in panel a. 

The measurement shown in Fig.\,\ref{supfig_Qis_Bz} emphasizes the circuit's pronounced susceptibility to out-of-plane magnetic fields. Interestingly, the maximum of the qubit frequency does not necessarily coincide with the maximal internal quality factor. The criterium for the in-plane field alignment is maximizing the qubit frequency versus $B_z$. By performing similar sweeps of $B_z$ for different values of $B_y$, we estimate the misalignment between the HH field and the sample's in-plane direction to be $0.7^\circ$.

\section{Qubit response to in-plane magnetic fields}
\label{ASEC:Magnetic_field_dependence}
The in-plane magnetic field dependence of the qubit transition frequency $f_1 (B)$ is calculated by mapping the qubit onto a linearized, lumped-element circuit model consisting of three inductive contributions in series, which are shunted by a capacitance $C_\mathrm{s} = 137\,\mathrm{fF}$. The inductive contributions arise from a field-independenent geometric inductance $L_\mathrm{s} = 0.45\,\mathrm{nH}$, and two field-dependent kinetic inductances associated with the pure Al and grAl thin films, $L_\mathrm{k,Al} (B)$ and $L_\mathrm{k,grAL} (B)$, respectively. The lumped-element model transition frequency is
\begin{equation}
f_1 = \frac{1}{2 \pi \sqrt{C_\mathrm{s} \left( L_\mathrm{k,Al}(B) +  L_\mathrm{k,grAl}(B) + L_\mathrm{s}\right)}}.
\label{EQ:f0_Field_dependence}
\end{equation}
The field dependence of the kinetic contributions is derived from Mattis-Bardeen theory for superconductors in the dirty limit \cite{MattisBardeen58} \begin{equation}
L_\mathrm{k} = \frac{\hbar R_\mathrm{n}}{\pi \Delta (B,T)},
\label{EQ:kinetic_inductance}
\end{equation}
where $R_\mathrm{n}$ is the normal state resistance and $\Delta (B,T)$ is the magnetic field and temperature dependent gap parameter. The dependence of the gap parameter on magnetic fields applied in-plane is derived from a two-fluid model (cf. p. 392 and 393 in \cite{Tin04})
\begin{equation}
\Delta (B, T = 0) / \Delta_{00} = \sqrt{\frac{1 - (B / B_\mathrm{c})^2}{1 + (B / B_\mathrm{c})^2}},
\label{EQ:Delta_B}
\end{equation}
where $\Delta_{00}$ is the gap parameter at zero temperature and zero magnetic field, and $B_\mathrm{c}$ is the critical magnetic flux density above which the pair correlation is zero. Due to the fact that the critical magnetic flux density of grAl \cite{Cohen68} is two orders of magnitude higher than that our pure Al films \cite{Meservey71}, we consider $L_\mathrm{k,grAl} (B)$ to be constant. However, for the Al kinetic inductance, by inserting Eq.\,\ref{EQ:Delta_B} into Eq.\,\ref{EQ:kinetic_inductance}, we find
\begin{equation}
L_\mathrm{k,Al} (B) = \underbrace{\frac{\hbar R_\mathrm{n}}{\pi \Delta_{00,\mathrm{Al}}}}_{L_\mathrm{k,Al}} \sqrt{\frac{1 + (B / B_\mathrm{c,Al})^2}{1 - (B / B_\mathrm{c,Al})^2}},
\label{AEQ:Lkin_field}
\end{equation}
where $L_\mathrm{k,Al}$ is the kinetic inductance in zero field. Using Eq.\,\ref{EQ:f0_Field_dependence} we fit the data presented in Fig.\,\ref{fig_field_dependence}a in the main text, and we obtain $L_\mathrm{k,Al} = 200\pm 5\,\mathrm{pH}$ and $B_\mathrm{c,Al} = 150\pm 5 \, \mathrm{mT}$.

The $200\,\mathrm{pH}$ value of $L_\mathrm{k,Al}$ corresponds to the intrinsic kinetic inductance of the Al film plus the contribution of the two contact areas between the grAl inductor and the Al electrodes. The presence of contact junctions is expected because the Al grains in grAl are uniformly covered by an amorphous $\mathrm{AlO}_x$ oxide. Taking into account the typically measured $15\%$ kinetic inductance fraction in Al thin films of comparable geometry \cite{Gruenhaupt17}, we estimate the intrinsic kinetic contribution of the Al film to be $70\,\mathrm{pH}$. The remaining $130\,\mathrm{pH}$ are associated with the contact junctions, from which we calculate a critical current $I_\mathrm{c} \approx 5\,\si{\micro\ampere}$ for each of them, with corresponding critical current density $j_\mathrm{c, JJ} \approx 0.13 \, \mathrm{mA / \si{\micro\metre}^2}$, comparable to the critical current density of the grAl film $j_\mathrm{c} = 0.4\, \mathrm{mA / \si{\micro\metre}^2}$ (cf. Eq.\,\ref{AEQ:critical_current_desnity}). The inductance participation of one contact junction in the total inductance is $p_{1, \mathrm{J}} = 2\%$. Using the energy participation ratio method presented in Ref.\,\cite{Minev19}, the nonlinear contribution per junction is
\begin{equation}
K_{1J} = \frac{h \omega_1^2}{4 \Phi_0 I_\mathrm{c}} p_{1J}^2 \approx 2 \pi \times 5\,\mathrm{kHz},
\end{equation}
where $\Phi_0 = h / 2 e$ is the magnetic flux quantum.

We calculate the kinetic inductance of the grAl film $L_\mathrm{k,grAl} = 2.7\,\mathrm{nH}$ by substracting $L_\mathrm{k,Al} = 200\,\mathrm{pH}$ (fitted, cf. Eq.\,\ref{AEQ:Lkin_field}) and $L_\mathrm{s} = 450\,\mathrm{pH}$ (FEM simulations, cf. App.\,\ref{ASEC:FEM_Simulation}) from the total inductance $L = 3.35\,\mathrm{nH}$, which we obtained from the measured resonance frequency $f_1 = 7.4887\,\mathrm{GHz}$ (cf. Fig.\,\ref{fig_fluorescence}) using $C_\mathrm{s} = 137\,\mathrm{fF}$ (FEM simulations, cf. App.\,\ref{ASEC:FEM_Simulation}). Considering the fact that the grAl volume consists of approximately 2.5 squares, the corresponding grAl sheet kinetic inductance is ${L_\mathrm{k, \square} = 1.1\,\mathrm{nH / \square}}$. This agrees with the value calculated from the room temperature sheet resistance $R_{\mathrm{n,\square}} \approx 1800\pm 200\,\si{\ohm} / \square$ using the Mattis-Bardeen theory for superconductors in the local and dirty limit \cite{MattisBardeen58}
\begin{equation}
L_\mathrm{K,\square}  = \frac{h R_{\mathrm{n},\square}}{2 \pi^2 \Delta_\mathrm{BCS}} \approx 1.3 \pm 0.1\,\mathrm{nH} / \square.
\end{equation}

\begin{figure}[!t]
\begin{center}
\includegraphics[width = \columnwidth]{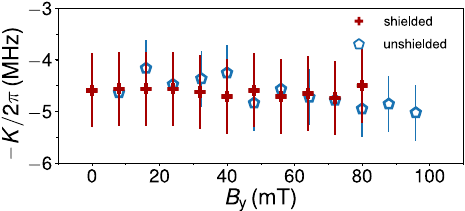}
\caption{\textbf{Measured non-linear coefficients K(5) (shielded) and K(6) (unshielded) versus in-plane field.}  }
\label{supfig_Kerr}
\end{center}
\end{figure}

As discussed in the main text, the grAl non-linearity $\alpha$ is consistent with a number $N \approx 6$ of effective JJs, yielding a critical current density 
\begin{equation}
j_\mathrm{c} = \frac{\Phi_0}{2 \pi} \frac{N}{L_\mathrm{k,grAl} A_\mathrm{grAl}} \approx 0.4 \, \mathrm{mA / \si{\micro\metre}^2},
\label{AEQ:critical_current_desnity}
\end{equation}
where $A_\mathrm{grAl} = 10 \times 200 \,\mathrm{nm^2}$ is the grAl cross section area. This value for $j_\mathrm{c}$ is in agreement with switching current measurements of similar grAl films \cite{Friedrich19}.

In Fig.\,\ref{supfig_Kerr} we show the field dependence of the non-linear frequency shift $K(5)$ and $K(6)$ measured during the same two cooldowns as the data presented in Fig.\,\ref{fig_field_dependence} (main text), with and without an outer Al shield, respectively. We chose to measure the non-linear coefficient for $n > 1$ in order to use a larger readout power and reduce the averaging time during the field sweep. As expected, since the ratio $\omega_1^2 / j_\mathrm{c}$ in the expression for $K$ remains constant, we do not observe a change in $K$ within the measurement accuracy.

\begin{figure}[!t]
\begin{center}
\includegraphics[width = \columnwidth]{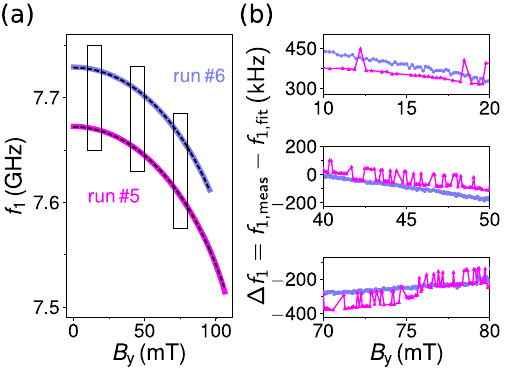}
\caption{\textbf{Interactions with other mesoscopic systems. a)} Qubit transition frequency $f_1$ as a function of the applied in-plane magnetic field $B_y$ measured in two different cool downs. The black dashed lines indicate fits to Eq.\,\ref{EQ:f0_Field_dependence}. \textbf{b)} Frequency difference $\Delta f_1$ between the measured transition frequency $f_{1, \mathrm{meas}}$ and the fit result $f_{1,\mathrm{fit}}$ (see Eq.\,\ref{EQ:f0_Field_dependence}) in three different field ranges. On this scale, discrete jumps in the qubit frequency become visible for the data taken in run \#5. The jumps remain qualitatively unchanged over a $150\,\mathrm{MHz}$ span covered by the qubit frequency versus $B_y$, and in neither of the runs did we observe signatures of avoided level crossings.}
\label{supfig_TLS}
\end{center}
\end{figure}

\begin{table*}[t]
\caption{\textbf{Summary of circuit parameters from all measurement runs:} qubit frequency $f_1$, external quality factor (zero-field) $Q_\mathrm{c,0}$, internal quality factor (zero-field) $Q_\mathrm{i,0}$, external coupling rate (zero-field) $\kappa_0$, internal decay rate (zero-field) $\gamma_0$, outer shielding configuration.}
\begin{center}
\begin{tabular}{||c || c c c c c c c ||} 
\hline
 & run \#1 & run \#2 & run \#3 & run \#4 & run \#5 & run \#6 & run \#7 \\ \hline \hline
$f_1\,\mathrm{(GHz)}$ & 7.6790 & 7.4887 & 7.5156 & 7.4778 & 7.672 & 7.7292 & 7.4749\\ 
$\alpha / 2 \pi\,\mathrm{(MHz)}$ & $3.90$ & $4.48$ & - & - & - & 3.35 & 4.5 \\
$Q_\mathrm{c,0} (\times 10^5)$ & $1.7 $ & $1.9 $ & $1.8 $ & $1.5 $ & $1.9$ & $1.9$ & $1.5$ \\
$Q_\mathrm{i,0} (\times 10^3)$ & $900$ & $750$ &  $95$ & $100$ & $130$ & $51$ & $750$\\
$\kappa_0 / 2 \pi\,\mathrm{(kHz)}$ & $45$ & $40$ & $42$ & $50$ & $40$ & $40$ & 51\\
$\gamma_0 / 2 \pi \,\mathrm{(kHz)}$ & $8.5$ & $10$ & $78$ & $75$ & $58$ & $150$ & 10\\
Outer shield: & & & & & & &\\
Cu & yes & yes & yes & yes & yes & yes & yes\\
Al & yes & yes & no & yes & no & no & yes\\
$\mu$-metall & yes & yes & no & no & no & no & yes \\
Magnetic field & no & no & yes & yes & yes & yes & no \\
\hline
\end{tabular}
\end{center}
\label{tab:Overview}
\end{table*}

\section{Interactions with other mesoscopic systems}
\label{ASEC:TLS}
Unwanted interactions between superconducting qubits and other mesoscopic systems are commonly observed \cite{Klimov18,LeSueur18,Schloer19,Lisenfeld2019}. Although the nature of the mesoscopic system is difficult to identify, and depends on the architecture of the qubit in general, among the most prominent suspects are defects present in the non-stoichiometric oxide of the JJ barrier and other interfaces \cite{Mueller19,Lisenfeld2019}, adsorbates and organic residuals from the fabrication process \cite{deGraaf18}, non-equilibrium quasiparticles \cite{deVisser11,Riste13,Pop14,Wang14}, and magnetic vortices \cite{Song09}. The interaction can be either transversal, causing a change of the underlying eigenbase due to a hybridization between the two systems, or longitudinal, inducing a frequency shift of the qubit that depends on the state of the mesoscopic system. For a true longitudinal coupling, the state dependent frequency shift does not depend on the frequency detuning between the systems. In the transversal case, the hybridization becomes visible in form of an avoided level crossing, when the two systems are tuned on resonance \cite{Martinis05}. 

We can tune the grAl transmon qubit transition frequency by more than $150\,\mathrm{MHz}$ by applying an external magnetic field in-plane. Figure\,\ref{supfig_TLS}a shows the qubit transition frequency as a function of the in-plane magnetic field $B_y$ measured in two consecutive cool downs (run\,\#5 and run\,\#6), as well as numerical fits according to Eq.\,\ref{EQ:f0_Field_dependence} (black dashed lines). Each frequency value is extracted from a fit to the resonance fluorescence response of the qubit similar to Fig.\,\ref{fig_fluorescence}. In both measurements shown, and in any other measurement of the same kind, we do not observe avoided level crossings. Figure\,\ref{supfig_TLS}b depicts the frequency difference $\Delta f_1$ between the measured transition frequency $f_{1,\mathrm{meas}}$ and the fitted transition frequency $f_{1,\mathrm{fit}}$, in three different field ranges. In run\,\#5, the qubit frequency is observed to jump between two metastable states, while it shows no significant discontinuity in run\,\#6. The fact that the observed frequency difference between the two metastable states during run\,\#5 is independent of the transition frequency of the qubit, suggests that the coupling is longitudinal.

\begin{figure}
\begin{center}
\includegraphics[width = \columnwidth]{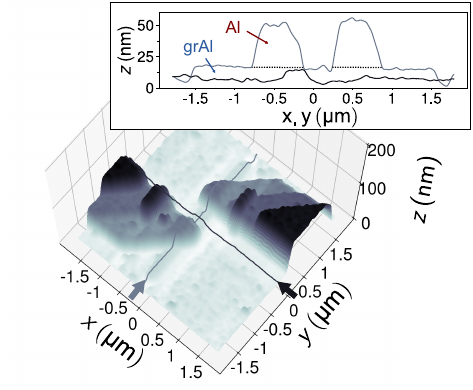}
\caption{\textbf{Film height profile.} Atomic force microscope image of the film height $z$ in the area around the grAl inductor, measured on a sample fabricated on the same wafer as the sample presented in the main text. The inset shows the cross section along $x$ (black) and $y$ (grey), as indicated by the arrows and overlay lines in the 3D plot. The cross section along the grAl inductor (grey) confirms the $t_\mathrm{grAl} = 10\,\mathrm{nm}$ grAl thickness and $t_\mathrm{Al} = 40\,\mathrm{nm}$ for each Al layer.}
\label{supfig_AFM}
\end{center}
\end{figure}

\begin{table}[b]
\caption{\textbf{Metal deposition parameters:} film thickness $t$, evaporation angle $\beta$, deposition rate $r$, absolute pressure in deposition chamber $p$.}
\begin{center}
\begin{tabular}{|| c || c c c c ||} 
\hline
layer & $t\,\mathrm{(nm)}$  & $\beta\,\mathrm{(^\circ)}$ & $r\,\mathrm{(nm/s)}$ & $p\,\mathrm{(mbar)}$ \\ \hline \hline 
grAl & 10 & 0 & 0.3 & $\sim 10^{-5}$ \\
Al & 40 & $+35$ & 1.0 & $\sim 6 \times 10^{-8}$ \\
Al & 40 & $-35$ & 1.0 & $\sim 6 \times 10^{-8}$ \\ \hline
\end{tabular}
\end{center}
\label{tab:fabrication}
\end{table} 

\vspace{20px}

\section{Summary of circuit parameters in all cooldowns}
\label{ASEC:sample_overview}
In Tab.\,\ref{tab:Overview} we summarize the circuit parameters measured in each of the four cooldowns. The time intervals between subsequent runs, during which the sample is at room temperature and atmospheric pressure, therefore subjected to aging, are the following: run \#1 to \#2 - 20 days, run \#2 to \#3 - 50 days, run \#3 to \#4 - 20 days, run \#4 to \#5 -  176 days, run \#5 to \#6 -  3 days, run \#6 to \#7 - 102 days.

\section{Film height profile}
\label{ASEC:AFM}
Figure\,\ref{supfig_AFM} shows an atomic force microscopy (AFM) image of the film height around the grAl volume for a sample fabricated in the same batch. The height profile shows several steps which originate from the three-angle evaporation process with angles $0^\circ$ (grAl, $t_\mathrm{grAl} = 10\,\mathrm{nm}$) and $\pm35^\circ$ (Al, $t_\mathrm{Al} = 40\,\mathrm{nm}$ each). The metal deposition parameters are summarized in Tab.\,\ref{tab:fabrication}. The main features are the leads connecting the grAl volume. Close to the edge of the image, all three layers overlap and the total film height is around $90\,\mathrm{nm}$. The inset in Fig.\,\ref{supfig_AFM} shows a cross section along the short (black) and the long (grey) edge of the grAl volume. The AFM measurement confirms the grAl film thickness of $t_\mathrm{grAl} = 10\,\mathrm{nm}$.

\end{document}